\documentclass[12pt]{article}

\usepackage{graphicx,amsmath,amsthm,amssymb,array,latexsym,cite,epsfig,finalT}

\numberwithin{equation}{section}

\input hmeas.def

\begin{document}

%\begin{spacing}{1}
\title{Hidden measurements, hidden variables and the volume
representation of transition probabilities
%\thanks{\emph{PACS No:} 03.65.Ta}
}
\author{
Todd A. Oliynyk
\thanks{Present address:
Max-Planck-Institut f\"{u}r Gravitationsphysik, Albert-Einstein-Institut,
Am M\"{u}hlenberg 1, D-14476 Golm, Germany$\;\;$ Email: todd.oliynyk@aei.mpg.de}\\
Department of Mathematical and Statistical Sciences\\
University of Alberta\\ Edmonton AB T6G 2G1
\\ Email: oliynyk@math.ualberta.ca}
\date{}

\maketitle

\vspace{2.0cm}
%\begin{spacing}{2}
We construct, for any finite dimension $n$,
a new hidden measurement model for
quantum mechanics based on representing quantum
transition probabilities by the volume of
regions in projective Hilbert space. For $n=2$
our model is equivalent to the Aerts sphere model
and serves as a generalization of it for dimensions
$n \geq 3$. We also show how to construct a hidden variables scheme
based on hidden measurements
and we discuss how joint distributions arise in our
hidden variables scheme and their relationship with
the results of Fine \cite{Fine}.
%\end{spacing}
\vspace{1.0cm}


\newpage

%\end{spacing}{1}

%\begin{spacing}{2}
\sect{intro}{INTRODUCTION}

Hidden measurements were introduced by Aerts \cite{Aert86} to
show that it is possible
to understand quantum probabilities as arising
from a lack of knowledge about the interactions
between a measuring device and the system that
it is measuring. In this way, 
quantum mechanics can be understood within 
classical probability theory with the peculiarities
of quantum probabilities arising from a
simple lack of knowledge and not some
mysterious source. 
The hidden measurement formalism
is just one of a number of approaches that try
to find a classical representation for
probability structures in quantum mechanics.
For example, stochastic extensions of
the Schr\"{o}dinger equation have been
proposed to account for the collapse
of the wave function \cite{ABBH,GP} and hence the
appearance of quantum probabilities. 
It has also been observed that quantum like behavior 
can arise within classical systems 
\cite{Pit,Cza92,Aert86,ACHV97,Kirk03a,Kirk03b}. 
The fact that it is possible to find classical
representations for the probability structures in quantum
mechanics and that quantum like behavior arises in non-quantum
systems hints that eventually quantum mechanics will
be understood within classical probability theory.

To describe hidden measurements, we denote by $X$ the
set of possible states of a system $S$ and by $Y$
the set of possible states of the measuring
device $\Mc$ used to measure $S$. Given that the
system is in a state $x\in X$ and the measuring
device is in the state $y\in Y$, we let
$z = \Mc(x,y)$ denote the outcome of the measurement.
The measurement is assumed to be deterministic so
if we let $Z$ denote the collection of all possible
measurement outcomes then 
$(x,y) \mapsto \Mc(x,y)$ defines a map
from $X\times Y$ to $Z$.
The fact that the result of a measurement
depends on the state of the measuring device is
the justification for the name ``hidden measurements''.

To formalize the above discussion, we need two measure
spaces $(X\times Y,\Ac(X\times Y))$ and $(Z,\Ac(Z))$.
Here we are using $\Ac(X\times Y)$
and $\Ac(Z)$ to denote $\sigma$-algebras on $X\times Y$
and $Z$, respectively. A \emph{deterministic measurement} 
is defined to be a measurable map (i.e. a random variable)
\leqn{measureA}{
\Mc : X\times Y \longrightarrow Z \, .
}
The state of the system plus measuring device is assumed
to be uncertain and characterized by a probability measure 
$\mu$  on  $(X\times Y,\Ac(X\times Y))$. The
probability of a measurement yielding an event $U\in \Ac(Z)$
is then given by
\leqn{measureB}{
\Prob(\Mc \in  U \, |\, \mu) := \int_{\Mc^{-1}(U)}\mu \, .
}
From this it can be seen that the probability of a measurement
obtaining a certain value is due not only to the uncertainty
of the system but also to the uncertainty in
the state of the measuring device which is characterized
by the measure $\mu$. Even in the situation where there
is no uncertainty in the state of the system there
can still be uncertainty in the state of the measuring
device and hence the outcome of a measurement is 
probabilistic. 
In the terminology of \cite{SAerts},
a deterministic measurement is referred to as
\emph{rule of interaction}. In
that paper, the special case 
where the
measure $\mu$ in \eqref{measureB}
factors as a product
of measure on $X$ and $Y$ (i.e. $\mu = \mu_{X}\mu_{Y}$) 
is called an \emph{interactive probability model}. 
For more discussion on the philosophical foundations
of the hidden measurement formalism we refer the
readers to the papers \cite{Aert86,Aert98} and
references cited therein.

The hidden measurement formalism has 
continued to be developed by a 
number of authors and various hidden measurement schemes
have been constructed \cite{Coec95a,Coec95b,Coec95c,ACHV97,Coec98a,
Coec98b,SAerts}.  The range of possible
hidden measurement schemes have been classified in
\cite{Coec97a,Coec97b} and for finite dimensional
quantum systems no preferred
scheme is identified. Criteria for selecting
out a preferred scheme is still lacking 
and thus different hidden measurement 
schemes should be investigated so
their relative merits can be compared.

The aim of this article is to introduce a new hidden
measurement scheme for finite dimensional quantum systems
based on the concept of representing
quantum transition probabilities by the volume of regions
of projective Hilbert space. Since
the measure we use in constructing the hidden
measurement scheme factors, our construction
defines an interactive probability model. 
For dimension $n=2$
our scheme is isomorphic to
the sphere model of Aerts \cite{Aert86,Cza92}. This can be
seen through the isomorphism $\PH \cong S^{2}$. 
For dimensions $n \geq 3$ our
approach offers a significant improvement over previous schemes
in that it is geometrical in origin and is formulated
on projective Hilbert space (i.e. we take
$X=Y=\PH$ for some Hilbert space $\PH$ ) which is the 
natural state space of quantum mechanics.
This allows for
a clear understanding of how the group of unitary transformations
acts on our hidden measurement scheme. 

With the exception of the
$2$ and $3$ dimensional models presented
in \cite{Aert86} and \cite{ACHV97},
previous hidden measurement models
have essentially used, in the notation above,
$X = \Hc$ or $S(\Hc)$ and for $Y$ a simplex sitting in $\Rbb^{n}$.
The choice of the simplex for $Y$ arose from the observation
that given a quantum state $\psi$ and an orthonormal
basis $\{\psi_{1},\ldots,\psi_{n}\}$, the transition probabilities
$p_{\psi_{k}}(\psi) := |\ip{\psi}{\psi_{k}}|^2$ regarded
as a vector $(p_{\psi_{1}}(\psi),\ldots,p_{\psi_{n}}(\psi))$
must lie in the simplex $\{ (x_{1},\ldots,x_{n}) \,|\,
x_{i}\geq 0\; \text{and} \; \sum_{i}x_{i} = 1\}$. The
hidden measurement scheme is then implemented by
partitioning the simplex into $n$ regions with volume
equal to the transition probabilities $p_{\psi_{k}}(\psi)$. 
The limitations of the scheme is that for each commuting
set of observables a new simplex must be introduced
and a priori it is unclear how the different measurement systems are
related. We note that
in \cite{Coec98c} it is shown that
by fixing one hidden measurement scheme
for one set of commuting observables that
it is possible to use a group translation
procedure to induce hidden measurement schemes
for the other commuting sets of observables
in a manner that is consistent with
quantum mechanics. %(see Proposition 1).
However, due to the abstract method of enforcing
the action of the unitary group it is difficult
to get a global picture of the measurement scheme
and the relations between different observables.
In contrast our method 
supplies a hidden measurement scheme for each
set of commuting observables and at the same
time provides a natural action of 
the unitary group on the schemes and shows
that the entire collection is compatible with
quantum mechanics. The action of the unitary group
is natural and easy to understand.

We also show how to construct a hidden variables scheme
based on hidden measurements. Here we are taking the term 
\emph{hidden variables} to mean representing quantum observables
and quantum states as random variables and probability distributions,
respectively, on a fixed space. Due to the nature of
the measurement depending on the state of the
measuring device, the type of hidden variables
that we construct are \emph{contextual}. A general
discussion of this point can be found in \cite{Coec98c}
where it is shown that for any hidden measurement system
it is possible to introduce a (non-unique) contextual hidden variables
theory. %(see Theorem 1).
Finally, we  discuss how joint 
distributions for commuting observables
arise in our hidden variables scheme and their relationship 
with the results of Fine \cite{Fine}.

\sect{PHS}{PROJECTIVE HILBERT SPACE}

In this section we review some basic results about projective Hilbert
space. We use the  book \cite{k6081} as our standard reference.
Let $(\Hc,\ip{\cdot}{\cdot})$ be a complex Hilbert space where the inner product
$\ip{\cdot}{\cdot}$ is taken to be linear in the second variable. Define 
\leqn{Hcross}{
\Hcx := \Hc\setminus\{0\} \quad \text{and} \quad \Cbbx := \Cbb\setminus\{0\}.
} 
On $\Hcx$ we can define an equivalence relation $\sim$ by 
$\psi \sim \phi$ if and only if there exists a $\lambda \in \Cbbx$ such that 
$\psi = \lambda \phi$.
Letting $[\psi]$ denote the equivalence class for $\psi \in \Hcx$, 
we have
\leqn{eclass}{
[\psi] := \{\, \lambda \psi \, |\, \lambda \in \Cbbx\,\} \; .
}
Projective Hilbert space $\PH$ is then defined as
\leqn{PH}{
\PH := \Hcx/\sim = \{\, [\psi] \,|\, \psi \in \Hcx\, \} \, .
}
It is well known that $\PH$ carries a 
Hilbert manifold structure for which the canonical projection
%\eqn{proj}{
$\pi : \Hcx \rightarrow \PH\; ; \;\psi \mapsto [\psi]$
%}
is a $C^{\infty}$ submersion. As a consequence for any $q\in \PH$ and
$v_{q} \in \T_{q} \PH$ there exists a $\psi \in \Hcx$ and $\phi \in \Hc$
such that
$q = [\psi]$ and $v_{q} = \T_{\psi} \pi\cdot \phi$. Here we
are using $T_{q}\PH$ to denote the tangent space of $\PH$
at $q\in \PH$ and $T_{\psi} \pi : \T_{\psi}\Hcx \cong \Hc \rightarrow
T_{[\psi]}\PH$ to denote
the tangent map of the mapping $\pi :\Hcx \rightarrow \PH$.
The above representations for points and tangent vectors on $\PH$
can be used to define a complex structure $\Jbb$ and a strongly
non-degenerate symplectic form $\omega$
on $\PH$ via the formulas
\leqn{cstruc}{
\Jbb(\T_{{\psi}} \pi \cdot \phi) := \T_{\psi}\pi \cdot i\phi 
}
and
\leqn{sympl}{
\omega_{[\psi]}(\T_{\psi}\pi\cdot\phi_{1},\T_{\psi}\pi\cdot\phi_{2}) := 
2\hbar\|\psi\|^{-4} \text{Im}\left( \ip{\phi_{1}}{\phi_{2}}\|\psi\|^{2}
-\ip{\phi_{1}}{\psi}\ip{\psi}{\phi_{2}}\right) 
}
for every $\psi \in \Hcx$ and $\phi,\phi_{1},\phi_{2}\in \Hc$. Recall
that a symplectic form is a non-degenerate closed two form. 
It should be noted that 
$g(v,w) = \omega(v,\Jbb w)$
defines a Riemannian metric on $\PH$ and hence 
establishes that $\PH$ is a K\"{a}hler manifold.

Given a function $f \in  C^{\infty}(\PH)$, the
non-degeneracy of the symplectic form $\omega$ implies
that the following equation
\leqn{Hamilton}{
\omega(X_{f},Y)  = \td f(Y) \quad \text{for all vector fields $Y$ on $\PH$},
}
uniquely defines a vector field $X_{f}$.
The Poisson bracket $\{\cdot,\cdot\}$ is then defined via
\leqn{poissonB}{
\{f,g\} := \omega(X_{f},X_{f}) \quad \forall \; f,g \in  C^{\infty}(\PH) \; .
}

Let $\Lc(\Hc)$ denote the set of bounded linear operators on $\Hc$.
Then the unitary group $\U$ is defined by
\leqn{group}{
\U := \{ U \in \Lc(\Hc) \, | \, \ip{U\psi}{U\phi} = \ip{\psi}{\phi} \quad
\forall \psi,\phi \in \Hc \, \} \; .
}
Its Lie algebra $\uf$ is the set of skew-adjoint operators, i.e.
\leqn{liealg}{
\uf := \{ A \in \Lc(\Hc) \, | \, A^{\dagger} = - 
A \, \} \, .
}
Here we are using $\dagger$ to denote the adjoint of an operator.
The following map
\leqn{gaction}{
\rho \; : \; \U \times \PH \rightarrow \PH \; : \; (U,[\psi]) \rightarrow
[U\psi] 
}
defines an action of $\U$ on $\PH$ by symplectomorphism (i.e. 
$\rho_{U}^{*}\omega = \omega$ for all $U\in \U$). There also exists
an equivariant momentum mapping $\J\, :\, \PH \longrightarrow \uf^{*}$ for this action defined by
\leqn{mmap}{
\langle \J([\psi]),A\rangle := 
i\hbar\frac{\ip{\psi}{A\psi}}{\|\psi\|^{2}} \quad
\forall \; \psi \in \Hc,\, A\in \uf \, , 
}
where $ \langle \cdot, \cdot \rangle $ denotes the canonical pairing between
$\uf^{*}$ and $\uf$. 
Letting $\mathcal{C}^{\infty}(\PH)$ denote
the set of smooth functions on $\PH$, the momentum map can be viewed
as a map
$\J\, :\, \uf \longrightarrow \mathcal{C}^{\infty}(\PH)$ by
defining
\leqn{mmapA}{
\J(A)(x) := \langle \J(x),A\rangle \quad \forall \; x \in \PH\; .
}
Recall that the defining property of a
momentum map is that
\leqn{emmap}{
\omega(\underline{A},Y) = \td \J(A)(Y) 
}
holds for all vector fields $Y$ on $\PH$ and 
$A\in \uf$ where $\underline{A}$ is the vector field on $\PH$ generated
by $A$, i.e. 
\leqn{generator}{
\underline{A}([\psi]) := \frac{d\,}{dt}\Bigl|_{t=0}
\rho_{\exp(tA)}([\psi]) = T_{\psi}\pi\cdot A\psi ,
}
while an equivariant momentum map satisfies the additional
condition
\leqn{equivariance}{
\langle\J\circ \rho_{U}(x),A\rangle = \langle J(x),U^{-1}AU
\rangle\quad \forall \, x\in \PH, A\in \uf , U\in \U \, .
}
It follows from the equivariance that
\leqn{poisson}{
\{\J(A),\J(B)\} = \J([A,B]) 
\quad \forall \; A,B \in \uf\, .
}

Let $\Sa$ denote the set of bounded self-adjoint operators
on $\Hc$. For each operator $H \in \Sa$ 
\leqn{exp1}{
\E{H} := \J(-\ioh H) \, 
}
defines a smooth function on $\PH$. This function
is just the usual expectation of the
observable $H$, i.e.
\leqn{exp2}{
\E{H}([\psi]) = \frac{\ip{\psi}{H\psi}}{\|\psi\|^2}\, . 
}
With this notation \eqref{poisson} can be written
in the more familiar form 
\leqn{poissonA}{
\{\E{A},\E{B}\} = \E{\ioh[B,A]} \; .
}

\sect{action}{ACTION ANGLE COORDINATES ON $\PH$}

For the remainder of this article, we will assume
that $\dim \Hc = N < \infty$. Let $\{\psi_{1},\ldots,\psi_{N}\}$
be an orthonormal basis for $\Hc$. Define the projection operators
\leqn{eqA}{
P_{\psi_{k}} = |\psi_{k}\rangle\langle\psi_{k}| \, .
}
We can use the momentum map to define smooth functions
$p_{\psi_{k}}$ on $\PH$ by
\leqn{eqB}{
p_{\psi} := \E{P_{\psi_{k}}} \, .
}
Using \eqref{exp2} we get that
\leqn{eqC}{
p_{\psi}([\psi]) = \frac{\ip{\psi_{k}}{\psi}}{\|\psi\|}^2
}
which is the transition probability from the state $\psi$ to
$\psi_{k}$. As the operators $P_{\psi_{k}}$ commute,
formula \eqref{poissonA} shows that the functions
$\{p_{\psi_{1}},\ldots,p_{\psi_{N}}\}$ are in involution, i.e.
\leqn{eqD}{
\{\psi_{j},\psi_{k}\} = 0 \quad 
\forall \; j,k =1,2,\ldots N \, .
}
It follows from $\id = \sum_{k=1}^{N} P_{\psi_{k}}$ that
\leqn{eqE}{
\sum_{j=1}^{N} p_{\psi_{k}} = 1\, ,
}
which shows that at most $(N-1)$ of the functions $p_{\psi_{k}}$
can be independent. It is not hard to show that the
set $\{p_{\psi_{2}},\ldots,p_{\psi_{N}}\}$ is independent.
That is the set of points in $\PH$ for which the
covectors $\{dp_{\psi_{2}},\ldots,dp_{\psi_{N}}\}$
are linearly dependent has measure zero. Consequently, we
can use these functions to construct action angle coordinates
of $\PH$ following the standard recipe, see \cite{k6081}
for details. This results in the following coordinate chart
\lalign{eqF}{
&\tau : T^{N-1}\times S \longrightarrow \PH \notag \\
&(\theta,\mathbf{I}) = ((\theta_{2},\ldots,
\theta_{N}),(I_{2},\ldots,I_{N}))\mapsto
\Bigl[ \bigl(1-\sum_{k=2}^{N}I_{k}\bigr)^{1/2}\psi_{1}
+ \sum_{j=2}^{N}e^{-i\theta_{j}}\sqrt{I_{j}}\psi_{j}
\Bigr] \label{eqF1}
}
where $T^{N-1}$ is the $(N-1)$ torus and
\leqn{eqG}{
S := \Bigl\{\, (I_{2},\ldots,I_{N})
\in \Rbb^{N-1} \, | \, 0 < I_{j}\,\; \sum_{j=2}^{N} I_{j} < 1 
\Bigr\} \, .
}
In this chart, the symplectic form $\omega$ is
given by
\leqn{eqH}{
\omega = \hbar \sum_{j=2}^{N} d\theta_{k} \wedge dI_{k} \, .
}
We also note that the functions $p_{\psi_{k}}$ have
the coordinate representations
$p_{\psi_{j}}(\theta,\mathbf{I}) = I_{j}$ for
$j = 2,3,\ldots, N$.
Using $\omega$, we can define a volume form $\nu$ on $\PH$ by
\leqn{eqJ}{
\nu := \left(\frac{-1}{2\pi\hbar}\right)^{N-1}\underset{\text{N-1 times}}{
\underbrace{
\omega\wedge \ldots \wedge \omega}} \; .
}
Locally this is given by
\leqn{eqK}{
\nu = \frac{(N-1)!}{(2\pi)^{(N-1)}}
\, d\theta_{2}\wedge\ldots\wedge\theta_{N}\wedge
dI_{2}\wedge\ldots\wedge dI_{N} \, .
}
Then because the chart \eqref{eqF1} covers all of $\PH$
except for a set of measure zero, the volume of
$\PH$ is given by
\leqn{eqL}{
\Vol(\PH) = \frac{(N-1)!}{(2\pi)^{(N-1)}}
\int_{0}^{2\pi}\ldots\int_{0}^{2\pi}
d\theta_{2}\ldots d\theta_{N} \int_{S}
dI_{2}\ldots dI_{N} \, .
}
A straightforward calculation shows that
%\eqn{eqM}{
$\int_{S} dI_{2}\ldots dI_{N} = 1/(N-1)!$
%}
and hence $\Vol(\PH) = 1$.

\sect{vreps}{VOLUME REPRESENTATION OF TRANSITION PROBABILITIES}

Suppose $\psi,\phi \in \Hcx$. Then the transition probability from
the state $\psi$ to $\phi$, or vice versa, is given by
\leqn{trans}{
T(\psi,\phi) := \frac{|\ip{\psi}{\phi}|^{2}}{\|\psi\|^{2}\|\phi\|^{2}}\;.
}
As this formula is invariant under scaling of $\phi$ or $\psi$ by
non-zero complex numbers, it passes to a well defined function on 
$\PH\times \PH$ given by
\leqn{transA}{
T([\psi],[\phi]) = \frac{|\ip{\psi}{\phi}|^{2}}{\|\psi\|^{2}\|\phi\|^{2}}\;
\quad \forall\; \psi,\phi \in \Hcx\, .
}
\begin{figure}
\begin{center}
\epsfig{file=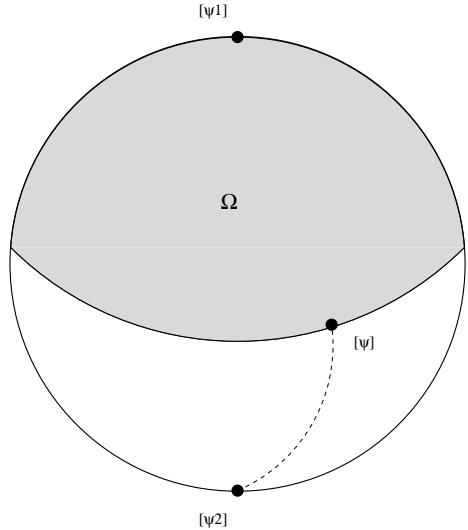,height=7cm}
\end{center}
\caption{Volume representation of transition probabilities} \label{fig1}
\end{figure}
It was shown in \cite{k6678} that if we let $d(x,y)$ denote the geodesic distance between points $x,y \in \PH$ then the distance $d(x,y)$ is related to
the transition probability $T(x,y)$ via the formula
\leqn{transC}{
T(x,y) = \cos^{2}\Bigl(\frac{d(x,y)}{\sqrt{2\hbar}}\Bigr) \, .
} 
This shows that there exists a representation of the transition probability
in terms of the geodesic distance. The question now is, are there other 
representations for the transition probability in terms of geometrical objects
on $\PH$? We will show that the transition probability, at least for finite
dimensional Hilbert spaces, can be related to the volume of certain regions
in $\PH$. To motivate this, we will first look at $\PH$ where $\Hc$ is
a 2 dimensional Hilbert space. Recall that $\PH \cong S^{2}$ where $S^{2}$ is
the ordinary two sphere in $\Rbb^{3}$. Suppose $\{\psi_{1},\psi_{2}\}$ is
an orthonormal basis for $\Hc$ and $\psi \in \Hcx$ is an  arbitrary state
vector. Since $\{\psi_{1},\psi_{2}\}$ is orthonormal, we can choose
them to be the north and south poles of $S^{2}$ as in Figure \ref{fig1}.
The symplectic form on $S^2$ is
\leqn{S2symp}{
\omega = \frac{\hbar \sin\theta}{2}d\phi \wedge d\theta
} 
while the volume form $\nu$ is given by
\leqn{S2vol}{
\nu = \frac{1}{2\hbar\pi}\omega \; .
}
The normalization on the volume form is chosen so that $\int_{\PH}\nu = 1$.
 
Referring to Figure \ref{fig1}, let $\Omega$ be the shaded region between
the points $[\psi_{1}]$ and $[\psi]$. Then a straightforward calculation shows
that
\leqn{volrepA}{
T([\psi_{2}],[\psi]) = \Vol(\Omega) := \int_{\Omega} \nu\, . 
}
Of course if we let $\gamma$ be the geodesic between $[\psi]$ and
$[\psi_{2}]$ represented by the dashed line in Figure \ref{fig1} then
we also have
\leqn{volrepB}{
T([\psi_{2}],[\psi]) = \cos^{2}\bigl((2\hbar)^{-1}
(\text{geodesic lenth of $\gamma$})\bigr)\, .
}
Letting $\Omega^{c}$ denote the complement of $\Omega$ we also have
\leqn{volrepC}{
T([\psi_{1}],[\psi]) = \Vol(\Omega^{c}) := \int_{\Omega^{c}} \nu\, . 
}
It is interesting to note that the conservation of probability
$1 = T([\psi_{1}],[\psi]) + T([\psi_{2}],[\psi])$
has the simple geometric representation
$\PH = \Omega \cup \Omega^{c}$.

To generalize the above construction to arbitrary but
finite dimensions we must first find a method for generalizing the
decomposition $\PH = \Omega\cup \Omega^{c}$. So for the moment,
let us still assume that $\dim \Hc = 2$ and that $\{\psi_{1},\psi_{2}\}$
is an orthonormal basis. Letting $\Omega_{1}:=\Omega$ and
$\Omega_{2} := \Omega^{c}$, a short calculation then shows that
\leqn{jointA}{
x \in \Omega_{k} \quad \text{if and only if} \quad p_{\psi_{j}}(x)
p_{\psi_{k}}(y) \geq p_{\psi_{k}}(x)p_{\psi_{j}}(y) \quad \text{for} \: 
j=1,2 \, ,
}
where $y=[\psi]$.  This motivates us to make the 
following definition. Let $\Hc$ be
an $N$ dimensional Hilbert space. Suppose $\beta = \{\psi_{1},\ldots,
\psi_{N}\}$ is an orthonormal basis for $\Hc$. Then define a region 
$\Omega(y,\beta,\psi_{j})$ of
$\PH$ that depends on a point $y\in \PH$, the basis $\beta$, and
a particular basis vector $\psi_{j}$ by
\leqn{Omega}{
\Omega(y,\beta,\psi_{k}) := \{\, x\in \PH\:|\: p_{\psi_{j}}(x)
p_{\psi_{k}}(y) \geq p_{\psi_{k}}(x)p_{\psi_{j}}(y) \quad 
\text{for} \: j=1,2,\ldots,N\,\}\, .
}

It is useful to introduce an alternate characterization for 
$\Omega(y,\beta,\psi_{k})$ which seems more
complicated but is actually easier to work with. To start, consider
the following vectors in $\Rbb^{N-1}$ 
\leqn{xivects}{
\xi_{1} := 0 \quad \text{and}\quad \xi_{j+1}:=(0,\ldots,\overset{\text{jth}}{1}
,\ldots,0) \quad j=1,2,\ldots,N-1\, .
}
Let
\leqn{xit}{
\xit := \sum_{j=1}^{N} p_{\psi_{j}}(y)\xi_{j} 
}
and define
\leqn{Sk}{
S(y,\beta,\psi_{k}) := \Bigl\{ \sum_{j=1,j\neq k}^{N} 
I_{j}\xi_{j} +
I_{k}\xit \in \Rbb^{N-1} \: \Bigl| \: I_{k}\geq 0 \quad \text{and} \quad \sum_{j=1}^{N} I_{j}=1
\Bigr\} \, .
}

\begin{prop} \label{jointC} \mnote{[jointC]}
Suppose $\beta = \{\psi_{1},\ldots,
\psi_{N}\}$ is an orthonormal basis for $\Hc$ and $y\in \PH$. Then
\leqn{jointC1}{
\Omega(y,\beta,\psi_{k}) = \Bigl\{ x \in \PH \:\Bigl| \:
\sum_{j=1}^{N} p_{\psi_{j}}(x)\xi_{j} \in S(y,\beta,\psi_{k}) \: \Bigr\} \, .
} 
\end{prop}
\begin{proof}
Assume that $p_{\psi_{k}}(y)\neq 0$. The case  $p_{\psi_{k}}(y) = 0$
will be left to the reader. Then
using \eqref{xit}, we can write $\sum_{j=1}^{N-1}p_{\psi_{j}}(x)\xi_{j}$  as
\leqn{jointC2}{
\sum_{j=1}^{N-1}p_{\psi_{j}}(x) = \sum_{j=1,j\neq k}^{N} \left( p_{\psi_{j}}(x)
-\frac{p_{\psi_{j}}(y)}{p_{\psi_{k}}(y)}p_{\psi_{k}}(x)\right)\xi_{j}
+ \frac{p_{\psi_{k}}(x)}{p_{\psi_{k}}(y)}\xit \, .
}
From \eqref{eqE} it is easy to see that
\lgath{jointC2}{
\sum_{j=1,j\neq k}^{N}\left( p_{\psi_{j}}(x)
-\frac{p_{\psi_{j}}(y)}{p_{\psi_{k}}(y)}p_{\psi_{k}}(x)\right) +
\frac{p_{\psi_{k}}(x)}{p_{\psi_{k}}(y)} = 1\, .
}
These two results along with $p_{\psi_{j}}\geq 0$ 
show that
\leqn{jointC3}{
\sum_{j=1}^{N-1}p_{\psi_{j}}(x)\xi_{j} \in S(y,\beta,\psi_{k})\quad \Longleftrightarrow \quad
p_{\psi_{j}}(x)p_{\psi_{k}}(y)
\geq p_{\psi_{j}}(y)p_{\psi_{k}}(x)
}
for $j=1,2,3,\ldots,N$. 
\end{proof}
The next two propositions show that the sets $\Omega(y,\beta,\psi_{j})$
have the required properties to be considered a generalization of the sets 
$\Omega_{1}=\Omega$ and $\Omega_{2}=\Omega^{c}$ from the previous section.

\begin{prop} \label{jointD} \mnote{[jointD]}
Suppose $\beta = \{\psi_{1},\ldots, 
\psi_{N}\}$ is an orthonormal basis for $\Hc$ and $y\in \PH$.
Then 
\leqn{jointD1}{
\PH = \bigcup_{k=1}^{N} \Omega(y,\beta,\psi_{k})
}
and
\leqn{jointD2}{
\emph{\Vol}\bigl(\Omega(y,\beta,\psi_{j})\cap \Omega(y,\beta,\psi_{k})\bigr)
 = 0 \quad
\text{for}\; j\neq k\, .
}
\end{prop}
\begin{proof}
Let $\Sbar$ denote the closure of $S$ defined by \eqref{eqG}, i.e.
\leqn{jointD3}{
\Sbar = \Bigl\{ (a_{2},\ldots,a_{N})\in 
\Rbb^{N-1}\: |\: a_{j}\geq 0 \; \text{and}\;
\sum_{j=2}^{N}a_{j} \leq 1 \; \Bigr\}
}
and define a map $\Jt :\PH \rightarrow \Rbb^{N-1}$ by
\leqn{jointD4}{
\Jt(x) = \sum_{j=2}^{N} p_{\psi_{j}}(x)\xi_{j} \, .
}
Since $p_{\psi_{j}}\geq 0$ and $\sum_{j=1}^{N}p_{\psi_{j}}=1$, we
have that $\Jt(\PH) \subset \Sbar$. To see that this inclusion is
actually an equality, consider any state vector 
$\phi = \sum_{j=1}^{N} c_{j}\psi_{j}$
where at least one of the coefficients $c_{j}$ is non-zero. Then
$\phi \in \Hcx$ and
\leqn{jointD6}{
\Jt([\phi]) := \frac{1}{\sum_{k=1}^{N}|c_{k}|^2}\sum_{j=2}^{N}|c_{j}|^2
\xi_{j}\, .
} 
It follows directly from this formula that $\Jt(\PH)=\Sbar$.
Also, it is not hard to verify that
\leqn{jointD7}{
\Sbar = \bigcup_{j=1}^{N} S(y,\beta,\psi_{j}) \, .
}
The above two results and proposition \ref{jointC} then imply that
$\PH = \bigcup_{j=1}^{N} \Omega(y,\beta,\psi_{j})$. 

From proposition \ref{jointC} and the definition of $\Jt$, we
have that $\Omega(y,\beta,\psi_{j}) = \Jt^{-1}(S(y,\beta,\psi_{j}))$.
Consequently 
\lalign{jointD8}{
\Omega(y,\beta,\psi_{j})\cap \Omega(y,\beta,\psi_{k}) 
&= \Jt^{-1}(S(y,\beta,\psi_{j})) \cap \Jt^{-1}(S(y,\beta,\psi_{k})) \notag \\
&= \Jt^{-1}(S(y,\beta,\psi_{j})\cap S(y,\beta,\psi_{k})) \, . 
}
But for $j\neq k$, the set 
$S(y,\beta,\psi_{j})\cap S(y,\beta,\psi_{k})$ lies inside an $N-2$
dimensional subset of $\Rbb^{N-1}$ and hence
$\Jt^{-1}(S(y,\beta,\psi_{j})\cap S(y,\beta,\psi_{k}))$ must
have measure zero. Therefore for $j\neq k$ the formula 
$\Vol\bigl(\Omega(y,\beta,\psi_{j})\cap \Omega(y,\beta,\psi_{k})\bigr)=0$
follows.
\end{proof}

\begin{prop} \label{jointE} \mnote{[jointE]}
Suppose $\{\psi_{1},\ldots,\psi_{N}\}$ is an orthonormal basis for $\Hc$ and $y\in \PH$.
Then
\leqn{jointE1}{
\emph{\Vol}(\Omega(y,\beta,\psi_{j})) = p_{\psi_{j}}(y) \qquad  j=1,2,\ldots,N \, .
}
\end{prop}
\begin{proof}
It is enough to prove it for $j=N$. From proposition \ref{jointC}, and
equations \eqref{eqH}-\eqref{eqK} it is clear that
\lalign{jointE2}{
\Vol(\Omega(y,\beta,\psi_{N})) &= \frac{(N-1)!}{(2\pi)^(N-1)}
\int_{0}^{2\pi}\ldots\int_{0}^{2\pi} d\theta_{2}\ldots
d\theta_{N} \int_{S(y,\beta,\psi_{N})}dI_{2}\ldots dI_{N} \notag\\
& = (N-1)! \int_{S(y,\beta,\psi_{N})}dI_{2}\ldots dI_{N} \label{jointE2.1}
}
But from \eqref{xit} and \eqref{Sk} it is easy to verify that
\leqn{jointE3}{
\int_{S(y,\beta,\psi_{N})}dI_{2}\ldots dI_{N} = \frac{1}{(N-1)!}
p_{\psi_{N}}(y) \, ,
}
which completes the proof.
\end{proof}

\sect{hm}{HIDDEN MEASUREMENTS}

We are now ready to construct our hidden measurment scheme.
To begin, let $A$ be a self-adjoint operator with spectral resolution
\leqn{spectA}{
A := \sum_{I=1}^{M}a_{I}P_{I}
}
where the projection operators $P_{I}$ can be further decomposed
into
\leqn{spectB}{
P_{I} := \sum_{j=1}^{m_{I}}
| \psi_{j,m_{I}}\rangle\langle \psi_{j,m_{I} }| \,
}
for some orthonormal basis $\{\, \psi_{j,m_{I}}\, | \, 1\leq I \leq M,
\;\; 1\leq j \leq m_{I}\}$.
Here $\{a_{1},\ldots,a_{M}\}$ are the distinct eigenvalues
of $A$ with multiplicities $\{m_{1},\ldots,m_{M}\}$. Note that
$n= m_{1} + \cdots + m_{M}$.

Let $\Hc_{I} := P_{I}(\Hc)$. Then each of the projection operators
defines a map
\leqn{projmapA}{
\Pt_{I} : \PH \setminus \pi((\Hc_{I}^{\perp})^{\times}) \longrightarrow
\PH \; ; \; [\psi] \longmapsto [P_{I}\psi] \, .
}
For notational convenience we  extend the maps $\Pt_{I}$ 
to all of $\PH$ by defining
\leqn{projmapB}{
\Pt_{I} : \PH \longrightarrow
\PH \; ; \; x \longmapsto 
\begin{cases}
\Pt_{I}(x) & \text{ if $x \in  \PH \setminus \pi((\Hc_{I}^{\perp})^{\times})$} \\
x & \text{otherwise}
\end{cases}\, .
}
Note that this map is essentially the linear map $\psi \rightarrow
P_{I}\psi$ projected down to $\PH$ where we have taken care
of the case when $P_{I}\psi = 0$.

Let $\beta = \{\, \psi_{j,m_{I}}\, | \, 1\leq I \leq M,
\;\; 1\leq j \leq m_{I}\}$ and define
\leqn{projmap}{
\Omega(y,\beta,P_{I}) := \text{int}\left( \bigcup_{j=1}^{m_{I}}
\Omega(y,\beta,\psi_{j,m_{I}}) \right) \, .
}
Then it follows from theorems \ref{jointE} and
\ref{jointD} that
\lgath{volumeAandvolumeB}{
\Vol(\Omega(y,\beta,P_{I})) = \sum_{j=1}^{m_{I}}p_{\psi_{j,m_{I}}}(y)
= \E{P_{I}}(y)  \, , \; \; 
\PH = \text{cl}\left( \bigcup_{I=1}^{M} \Omega(y,\beta,P_{I})\right)\, ,
\label{volumeA} \\
\intertext{and}
\Vol\bigl(\Omega(y,\beta,P_{I})\cap \Omega(y,\beta,P_{J})\bigr)
= 0 \quad I\neq J \, . \label{volumeB}
}
But $\E{P_{I}}([\psi])=\frac{\ip{\psi}{P_{I}\psi}}{\|\psi\|^{2}}$,
and hence \eqref{volumeA} and \eqref{volumeB}
provide a \emph{volume representation of
the transition probability}.

We define a deterministic measurement associated to
$A$ by
\leqn{measureC}{
\Mc_{A} : \PH \times \PH \longrightarrow \PH
\; ; \; (x,y) \longmapsto 
\begin{cases}
\Pt_{I}(x) & \text{if $y\in \Omega(x,\beta,P_{I})$} \\
  x & \text{otherwise}
\end{cases} \, .
}
To reproduce quantum mechanics, for each quantum state $x\in \PH$
we define a measure on $\PH\times \PH$ by
\leqn{measureD}{
\mu_{x} = \delta_{x}\times\nu
}
where $\delta_{x}$ is the Dirac measure on $\PH$ with support
at $x$ and $\nu$ is the volume form \eqref{eqJ}.
This choice of measure can be interpreted as saying that
we are certain that the system is in the state $x$ but
the measuring device is characterized by a uniform
distribution over its state space. Initially, we have
maximum information about the state of the system but
minimum information about the state of the measuring device.

From the definition of our deterministic measurement \eqref{measureC},
it is easy to see from \eqref{measureB} that given the
state $\mu_{x}$ there are exactly $M$ possible outcomes 
$\{\Pt_{1}(x),\ldots,\Pt_{M}(x)\}$ with probabilities
\leqn{measureE}{
\Prob( \Mc_{A}= \Pt_{I}(x)\,| \mu_{x} ) = \E{P_{I}}(x) \quad \text{for $
I = 1,2,\ldots,M$.}
}
This exactly reproduces the projection
postulate and hence quantum probabilities. 

\sect{hv}{HIDDEN VARIABLES}

Hidden measurements are a special case of hidden variables.
In this section we will write the hidden measurement scheme
introduced in the previous section as an explicit hidden
variables scheme. To accomplish this, for each 
self-adjoint operator $A$ we define a random variable 
\leqn{randomA}{
f_{A} : \PH\times \PH \longrightarrow \Rbb
}
by
\leqn{randomB}{  
f_{A} := \E{A}\circ \Mc_{A} \, .
}
Then from \eqref{volumeA} and \eqref{measureC} it is clear that
$f_{A}$ can take on only
$M$ distinct values $\{a_{1},\ldots,a_{M}\}$
and that
\leqn{randomB1}{
f_{A}^{-1}(a_{I}) = \bigcup_{y\in\PH}\{y\}\times 
\Omega(y,\beta,P_{I}) \, .
}
Therefore
\leqn{measureF}{
\Prob( f_{A}= a_{I}\,| \mu_{x} ) =
\int_{f_{A}^{-1}(a_{I})} \mu_{x}
= \E{P_{I}}(x) \quad \text{for $
I = 1,2,\ldots,M$}
}
by \eqref{volumeA} and \eqref{randomB1}
This reproduces all single observable measurements
in quantum mechanics. Note in particular that
we have the identity
\leqn{randomC}{
\E{A}(x) = \int f_{A}\mu_{x}  \, .
}

To completely reproduce 
quantum mechanics we must also deal with the joint
distributions of commuting observables.
So suppose $A$ and $A'$ have the following spectral resolution
\leqn{spectD}{
A = \sum_{I=1}^{M}a_{I}P_{I}\quad \text{and} \quad 
A' = \sum_{I=1}^{M'}a_{I}'P_{I}'\, ,
}
Fixing a common basis of orthonormal eigenvectors
$\beta =  \{\psi_{1},\ldots,\psi_{n}\}$
for $A$ and $A'$, it is easy to see that
\leqn{volumeC}{
\Omega(y,\beta,P_{I})\cap \Omega(y,\beta,P_{J}') =
\Omega(y,\beta,P_{I}P_{J}')
}
From this result, the definitions of $\Mc_{A}$ and $\Mc_{A'}$,
and \eqref{randomB1} we get
\leqn{randomB2}{
f_{A}^{-1}(a_{I}) \cap f_{A'}^{-1}(a_{J}')
= \bigcup_{y\in\PH} {y}\times \Omega(y,\beta,P_{I}P_{J}') \, .
} 
From this and \eqref{volumeA} it follows that 
\leqn{randomB3}{
\int_{f_{A}^{-1}(a_{I})\cap f_{A'}^{-1}(a_{J}')} \mu_{x}
= \Vol\bigl(\Omega(y,\beta,P_{I}P_{J}')\bigr) =
\E{P_{I}P_{J}'}(x) \, .
}
Therefore
\leqn{randomB4}{
\Prob( f_{A}= a_{I}, f_{A'} = a_{J}'\,| \mu_{x} ) =
\int_{f_{A}^{-1}(a_{I})\cap f_{A'}^{-1}(a_{J}')} \mu_{x}
= \E{P_{I}P_{J}'}(x)
}
for $1 \leq I \leq M$ and $1\leq J \leq M'$.
The generalization to 3 or more commuting observables
is obvious. We see from \eqref{measureF}
and \eqref{randomB4} that all of quantum mechanics
can be reproduced by the random variables $f_{A}$
and the measures $\mu_{x}$.

It is also worthwhile to take some time and examine the
relationship between $f_{A}$, $f_{A'}$, and
$f_{AA'}$ for commuting operators $A$ and $A'$. We would
expect that the random variables  $f_{AA'}$ and $f_{A}f_{A'}$ 
should be equivalent. To see this we first note
that from \eqref{spectD} we have
\leqn{spectD1}{
AA' = \sum_{I=1}^{M}\sum_{J=1}^{M'}a_{I}a_{J}'P_{I}P_{J}'\, .
}
For simplicity we assume that the values $\{a_{I}a_{J}'\, |\,
1\leq I\leq M,\; 1\leq J\leq M'\}$
are distinct. The following results hold true even if
the values $a_{I}a_{J}$ are not distinct 
and will be left to the reader.
From this and the definitions of $\Mc_{A}$, $\Mc_{A'}$, 
$\Mc_{AA'}$, $f_{A}$, $f_{A'}$, and $f_{AA'}$ it follows that
\leqn{volumeD}{
f_{AA'} = f_{A}f_{A'} \quad \text{a.s. with respect to the measure $\mu_{x}$} 
}
and hence the random variables $f_{AA'}$ and $f_{A}f_{A'}$
are indeed equivalent. 

We also note that density matrices can be easily incorporated
into our formalism. To see this, again suppose $\{\psi_{1},\ldots,\psi_{n}\}$
is an orthonormal basis and that
\leqn{densA}{
\rho = \sum_{j=1}^{N}p_{j}|\psi_{j}\rangle\langle \psi_{j}|
}
is a density matrix (i.e. $p_{i}\geq 0$ and $\sum_{j=1}^{n}
p_{j} = 1$). Due to the uncertainty in the
state of the system we replace the Dirac measure in
\eqref{measureD} by $\sum_{j=1}^{n}p_{j}\delta_{x_{j}}$
where $x_{j} = [\psi_{j}]$. So if we let
\leqn{densB}{
\mu_{\rho} = \Bigl(\sum_{j=1}^{n}p_{j}\delta_{x_{j}}\Bigr)\times \nu\, ,
}
it follows from \eqref{measureF} and \eqref{densB} that
\leqn{densC}{
\Prob(f_{A} = a_{I}\, | \, \mu_{\rho})
= \sum_{j=1}^{n}p_{j}\Prob(f_{A} = a_{I}\, | \, \mu_{x_{j}})
= \sum_{j=1}^{n}p_{j}\E{P_{I}}(x_{j}) \, .
}
But
\leqn{densD}{
\sum_{j=1}^{n}p_{j}\E{P_{I}}(x_{j}) = \sum_{j=1}^{n}p_{j} 
\ip{\psi_{j}}{P_{I}\psi_{j}} = \text{Tr}(\rho P_{I}) \, ,
}
and hence it follows that
\leqn{densE}{
\Prob(f_{A} = a_{I}\, | \, \mu_{\rho}) = \text{Tr}(\rho P_{I})
\quad \text{for $I=1,2,\ldots M$}
}
which reproduces the density matrix formalism in quantum
mechanics.  One point worth mentioning is that if
the density matrix $\rho$ had another expansion
\leqn{densF}{
\rho = \sum_{j=1}^{N}q_{j}|\phi_{j}\rangle\langle \phi_{j}|
}
in terms of a different orthonormal basis $\{\phi_{1},\ldots,\phi_{n}\}$
then according to above prescription we would associate to $\rho$ the measure
\leqn{densG}{
\mu_{\rho}' = \Bigl(\sum_{j=1}^{n}q_{j}\delta_{y_{j}}\Bigr)\times \nu
}
where $q_{j} = [\phi_{j}]$. Obviously $\mu_{\rho}'=\mu_{\rho}$
if and only if $\{\phi_{1},\ldots,\phi_{n}\}$ and 
$\{\psi_{1},\ldots,\psi_{n}\}$  are identical bases. However,
$\mu_{\rho}'$ and $\mu_{\rho}$ carry identical statistical
information as far as quantum mechanics is concerned because
\leqn{densH}{
\Prob(f_{A} = a_{I}\, | \, \mu_{\rho}) = 
\Prob(f_{A} = a_{I}\, | \, \mu_{\rho}')= \text{Tr}(\rho P_{I})
\quad \text{for $I=1,2,\ldots M$}
}
for every self-adjoint operator $A$ by equation \eqref{densE}. 
Thus $\mu_{\rho}'$ and $\mu_{\rho}$
are equivalent measures from the quantum point of view.

\sect{jd}{JOINT DISTRIBUTIONS}

Let us now consider $\ell$ self-ajoint operators $A_{1},A_{2},\ldots,A_{\ell}$,
which may or may not commute. From the previous section we
can associate to each observable to a random variable $f_{A_{k}}$
$k=1,2,\ldots \ell$
such that
\leqn{jd1}{
\Prob( f_{A_{k}} = a^{k}_{I}|\mu_{x} ) = \E{P^{k}_{I}}(x)
\quad 1 \leq I \leq M_{k}
} 
where
\leqn{jd2}{
A_{k} = \sum_{I=1}^{M_{k}} a^{k}_{I} P^{k}_{I}
}
is the spectral resolution of $A_{k}$. We are free to define
a joint distribution for the random variables 
$f_{A_{1}},f_{A_{2}},\ldots,f_{A_{\ell}}$
by
\leqn{jd3}{
\Prob(f_{A_{1}} = a^{1}_{I_{1}},\ldots,
f_{A_{\ell}} = a^{\ell}_{I_{\ell}})
= \int_{\bigcap_{k=1}^{\ell} f^{-1}_{A_{k}}(a^{k}_{I_{k}})}
\mu_{x} \, 
}
for $ 1 \leq I_{k} \leq M_{k}$, $k=1,2,\ldots,\ell$.
It would then appear from \eqref{jd1} and \eqref{randomB}
the joint distribution \eqref{jd3} would  have marginals
that agree with all the possible quantum distributions. 
However, this contradicts the results of Fine \cite{Fine}
where he showed that if there exists a 
joint probability distribution
with marginals that agree with  the quantum probability distributions,
wherever those are defined, then the correlations must satisfy Bell's 
inequalities. We know that for certain choices of observables
and states that Bell's inequalities are violated. But our
joint distribution \eqref{jd3} was derived for an arbitrary
set of observables $A_{1},A_{2},\ldots,A_{\ell}$ and so we
seem to have a contradiction.

The resolution of the contradiction is that the random variables
$f_{A}$ do not only depend on the observable $A$ but also
on the basis of orthonormal eigenvectors $\beta$. The dependence on the
basis vectors $\beta$ is clear
from the definition \eqref{measureC} of the
measurement maps $\Mc_{A}$ and the definition of
the random variables \eqref{randomB}. Therefore to be more precise
we will use the notation $f_{(A,\beta)}$ to make
clear this dependence. In the case where $A$ has distinct
eigenvalues there is only one basis $\beta$ and hence
a unique observable is associated to $A$. In all other cases
where there are degenerate eigenvalues, there
will be a family of random variables associated
to $A$. This is particularly important in deriving
the joint probability formula \eqref{randomB4} for
commuting observables. During the derivation, we used the random variables
$f_{(A,\beta)}$ and  $f_{(A',\beta)}$. The important
point to understand is that we had to use a common 
eigenbasis $\beta$ for
both $A$ and $A'$ to get the correct answer.
On the other hand, the single observable distributions
\eqref{measureF} are independent of the particular
eigenbasis chosen. 

To understand how the choice of basis affect the
joint distribution suppose that we have
four self-adjoint operators $A_{1}$,$A_{2}$,$B_{1}$,$B_{2}$ 
such that $A_{i}$ commute with the
$B_{j}$. In other words, each of the four pairs of
operators $\{A_{1},B_{1}\}$, $\{A_{1},B_{2}\}$,
$\{A_{2},B_{1}\}$ and $\{A_{2},B_{2}\}$
are separately diagonalizable. To associate random variables
to these observables we need to fix bases of eigenvectors.
So we will let $(A_{1},\alpha_{1})$, $(A_{2},\alpha_{2})$,
$(B_{1},\beta_{1})$, $(B_{2},\beta_{2})$ denote
the operator-eigenbasis pairs. Let
\leqn{jd4}{
A_{i} = \sum_{I=1}^{M_{i}} a^{i}_{I}P^{A_{i}}_{I} \quad
i=1,2 \quad \text{and} \quad
B_{j} = \sum_{J=1}^{N_{j}} b^{j}_{I}P^{B_{j}}_{J} \quad
j=1,2 
}
be the spectral resolutions for the operators $A_{i}$ and $B_{j}$.
We can define a joint distribution for the 
random variables $f_{(A_{i},\alpha_{i})}$, $f_{(B_{j},\alpha_{j})}$
by
\leqn{jd5}{
\Prob(f_{(A_{i},\alpha_{i})}=a^{i}_{I_{i}}, 
f_{(B_{j},\beta_{j})} = b^{j}_{J_{j}}\,|\,\mu_{x})
= \int_{V(a^{i}_{I_{i}},b^{j}_{J_{j}})} 
\mu_{x}
}
for $1\leq I_{i} \leq M_{i}$, $1\leq J_{j} \leq N_{j}$, and $i,j=1,2$
where
\leqn{jd6}{
V(a^{i}_{I_{i}},b^{j}_{J_{j}}) = 
f_{(A_{1},\alpha_{1})}^{-1}(a^{1}_{I_{1}})
\cap f_{(A_{2},\alpha_{2})}^{-1}(a^{2}_{I_{2}})
\cap f_{(B_{1},\beta_{1})}^{-1}(b^{1}_{I_{1}})
\cap f_{(B_{2},\beta_{2})}^{-1}(b^{2}_{I_{2}}) \, .
}
For the moment, consider the single marginal 
arising from the joint distribution
\lalign{jd7a}{
\Prob(f_{A_{1},\alpha_{1}}=a^{1}_{I_{1}}\,|\,\mu_{x})
&= \sum_{I_{2}=1}^{M_{2}}\sum_{J_{1}=1}^{N_{1}}
\sum_{J_{2}=2}^{N_{2}}
\Prob(f_{(A_{i},\alpha_{i})}=a^{i}_{I_{i}},
f_{(B_{j},\beta_{j})} = b^{j}_{J_{j}}\,|\,\mu_{x}) \notag \\
&= \int_{f_{(A,\alpha_{1})}^{-1}(a^{1}_{I})} \mu_{x}\, .
\label{jd7}}
From \eqref{volumeA} and \eqref{randomB1}
it is clear that 
\leqn{jd8}{
\int_{f_{(A,\alpha_{1})}^{-1}(a^{1}_{I})} \mu_{x}
= \E{P^{A_{1}}_{I}}(x)
}
is independent of the eigenbasis $\alpha_{1}$. Therefore
the joint distribution \eqref{jd5} will yield the
correct single variable quantum distributions
independent of a particular choice of the eigenbasis
$\alpha_{i}$ and $\beta_{j}$. However, when 
trying to satisfy the two variable quantum
distributions which arise from the fact
that the pairs of operators ${A_{i},B_{j}}$ 
are separately diagonalizable is where conflict appears.
So now consider the two variable marginal
\lalign{jd9}{
\Prob(f_{(A_{1},\alpha_{1})}=a^{1}_{I_{1}},
f_{(B_{1},\beta_{1})}=b^{1}_{J_{1}}\,|&\,\mu_{x})
= \sum_{I_{2}=1}^{M_{2}}\sum_{J_{2}=2}^{N_{2}}
\Prob(f_{(A_{i},\alpha_{i})}=a^{i}_{I_{i}},
f_{(B_{j},\beta_{j})} = b^{j}_{J_{j}}\,|\,\mu_{x})\notag\\
&= \int_{f_{(A,\alpha_{1})}^{-1}(a^{1}_{I_{1}})
\cap f_{(B,\beta_{1})}^{-1}(b^{1}_{J_{1}})} \mu_{x} \label{jd9.1}
}
In order to apply equation \eqref{randomB3} to ensure
that the two variable marginal agrees with
the quantum one, we must first assume
that $\alpha_{1}=\beta_{1}$. Therefore it
follows that  
\leqn{jd9}{
\Prob(f_{(A_{1},\alpha_{1})}=a^{1}_{I_{1}},
f_{(B_{1},\beta_{1})}=b^{1}_{J_{1}}\,|\,\mu_{x})
= \E{P^{A_{1}}_{I_{1}}P^{B_{1}}_{J_{1}}}(x)
} 
provided $\alpha_{1}=\beta_{1}$.
From this we can conclude that the joint
distribution \eqref{jd5} will yield
the correct quantum variable distributions
corresponding the the 
set of separately commuting observables  
$\{A_{1},B_{1}\}$, $\{A_{1},B_{2}\}$, $\{A_{2},B_{1}\}$,
$\{A_{2},B_{2}\}$ provided
$\alpha_{1}=\beta_{1}$, $\alpha_{1}=\beta_{2}$,
$\alpha_{2}=\beta_{1}$, and $\alpha_{2}=\beta_{2}$.
In other words $\alpha_{1}=\alpha_{2}=\beta_{1}=\beta_{2}$
and hence all the operators $A_{1}$,$A_{2}$, $B_{1}$, and
$B_{2}$ must be simultaneously diagonalizable.
Thus there is no contradiction with Fine's results \cite{Fine}.

We can make the following conclusions:
\begin{itemize}
\item[(i)]
to each self-adjoint operator $A=\sum_{I=1}^{M}a_{I}P_{I}$ and each
orthonormal eigenbasis $\beta$ of $A$ we can assign a random
variable $f_{(A,\beta)}$ defined by \eqref{randomB}
such that 
\leqn{measureFa}{
\Prob( f_{(A,\beta)}= a_{I}\,| \mu_{x} ) =
\int_{f_{(A,\beta)}^{-1}(a_{I})} \mu_{x}
= \E{P_{I}}(x) \quad 1\leq I \leq M
}
for each $x\in \PH$, and
\item[(ii)]
if $A=\sum_{I=1}^{M}a_{I}P_{I}$, and $A'=\sum_{I=1}^{M'}a_{I}'P_{I}'$
are simultaneously diagonalizable and $\beta$ is
a common eigenbasis then
\leqn{randomB4a}{
\Prob( f_{(A,\beta)}= a_{I}, f_{(A',\beta)} = a_{J}'\,| \mu_{x} ) =
\int_{f_{(A,\beta)}^{-1}(a_{I})\cap f_{(A',\beta)}^{-1}(a_{J}')} \mu_{x}
= \E{P_{I}P_{J}'}(x)
}
for $1 \leq I \leq M$, $1\leq J \leq M'$ and each
$x\in \PH$.
\end{itemize}
Note that equation \eqref{measureFa} is independent of
the basis $\beta$ and that \eqref{randomB4a} can be
easily generalized to 3 or more commuting observables.

%If for each self adjoint operator $A$ we let $\mathcal{B}(A)$
%denote the set of orthonormal eigenbases for $A$, then  
%\eqref{measureFa} and \eqref{randomB4a} show that in order
%to reproduce quantum mechanics we must represent a
%single observable $A$ by the set of random variables
%$\{ f_{(A,\beta)}\, |\, \beta \in \mathcal{B}(A) \}$ and
%each state $x\in \PH$ by the probability distribution
%$\mu_{x}$. This shows that only operators
%with distinct eigenvalues have associated to
%them a unique random variable. 

\sect{conc}{CONCLUSION}

We have, for any finite dimension N, constructed
a hidden measurement model for quantum mechanics
based on representing quantum transition probabilities
by the volume of regions in projective Hilbert
space. The geometrical nature of our construction
allows for a clear understanding of the
action of the unitary group on the hidden
measurement scheme in contrast to previous constructions.
We also showed how to construct a contextual
hidden variables theory based  on our hidden measurement
scheme. 

While the hidden measurement formalism is an
interesting way of looking at the theory
of quantum measurements, the obvious weakness
is that there is no physical principle
behind constructing the deterministic
measurements (see equations \eqref{measureA} and \eqref{measureC})
which are supposed to represent 
the interaction of the quantum system with
real measuring devices. Since the results of
this paper and previous work show that it is
possible to have a consistent hidden measurement
scheme for quantum mechanics, the question now
is - can a realistic dynamical theory 
for the measuring device plus the system
being measured be constructed
which singles out a particular form of the
deterministic measurement?
If this can be done in a compelling
manner then it would represent a significant
advance in our understanding of the quantum
theory of measurement. Some work in this
direction is contained in \cite{GP} where
the authors consider a
model with deterministic dissipative dynamics
for the quantum measurement process. In dimension
two, the geometry of the model is exactly the
same as the Aerts' sphere model.

\section*{ACKNOWLEDGEMENTS} 
\noindent This work was partially supported
by the ARC grant A00105048 at the University of Canberra and
by the NSERC grants A8059 and 203614 at the University of Alberta.
I would also like to thank the referee for his useful
comments and criticisms.
%\end{spacing}

%\begin{spacing}{2}

\bibliography{hmeas}

%\end{spacing}
\end{document}